\documentclass[preprint,showpacs,amsmath,amssymb,latexsymb]{revtex4}
\usepackage{graphicx} 
\begin{document}
\draft
\title{Functionalized pentacene field-effect transistors with logic circuit applications}
\author{Jin Gyu Park$^1$, Relja Vasic$^1$, James S. Brooks$^1$, John E. Anthony$^2$}
\address{$^1$National High Magnetic Field Laboratory, Florida State University, Tallahassee, FL 32310}
\address{$^2$Department of Chemistry, Univ. of Kentucky, Lexington, KY 40506 }

\date{\today }

\begin{abstract}
Funtionalized pentacene,
6,13-bis(triisopropylsilylethynyl)pentacene (TIPS-pentacene),
field-effect transistors(FET's) were made by thermal evaporation
or solution deposition method and the mobility was measured as a
function of temperature and light power. The field-effect mobility
($\mu$$_{\rm FET}$) has a gate-voltage dependent activation
energy. A non-monotonic temperature dependence was observed at
high gate voltage (V$_G$ $<$ -30 V) with activation energy E$_a$
$\sim$ 60 - 170 meV,depending on the fabrication procedure. The
gate-voltage dependent mobility and non-monotonic temperature
dependence indicates that shallow traps play important role in the
transport of TIPS-pentacene films. The current in the saturation
regime as well as mobility increase upon light illumination and is
proportional to the light intensity, mainly due to the
photoconductive response. Transistors with submicron channel
length showed unsaturating current-voltage characteristics due to
the short channel effect. Realization of simple  circuits such as
NOT(inverter), NOR, and NAND logic gates are demonstrated for thin
film TIPS-pentacene transistors.

\end{abstract}

\maketitle




\newpage
\section{Introduction}

The use of organic material for the electronic devices has great
importance for future application due to its low cost and easy
fabrication procedures. Pentacene, one of most promising organic
material for device application, has been studied intensively in
polycrystalline\cite{Jackson} or single crystalline
structures\cite{Ramirez} due to its high room temperature hole
mobility more than 1 cm$^2$/Vs. And other polyacene single
crystals such as tetracene\cite{Morpurgo}, anthracene\cite{Park}
as well as thin film structures\cite{Jackson_tet} were also
investigated.

By attaching the side functional group, soluble and stable
functionalized pentacene can be made\cite{Anthony} and this make
its fabrication methods versitile and increase application for
devices. Transport measurement depending on its crystal
direction\cite{Brooks1} shows different conducting properties and
the persistent photoconductivity was already
reported.\cite{Brooks2} Field-effect transistor (FET) application
of these functionalized pentacene was shown\cite{Jackson2,Payne}
and they can get room temperature mobility up to 0.4 cm$^2$/Vs
with deposition under substrate heating and gate dielectric with
self-assembled monolayer. Temperature dependence of mobility is
key to elucidate its quality and transport mechanism. In the case
of field-effect mobility, band-like temperature dependence is very
rare and only observed at high temperature range in single
crystal\cite{Morpurgo,Park} and thin film\cite{Jackson3}.

In addition to the electrical applications, its photosensitive
properties\cite{Brooks2} can be used as light detector or optical
switch. Light responsive properties of organic transistor have not
been investigated as much as electrical properties. The
performance of photoresponsivity in polymer\cite{Kumar,Noh} and
pentacene\cite{Liang} is still lower than that of amorphous
silicon ($\sim$ 300 A/W). However, this can be improved by
optimization of device fabrication

In this article, we report the electrical and optical properties
of 6,13-bis(triisopropylsilylethynyl)pentacene (TIPS-pentacene)
FET. For the electrical properties, we measured field-effect
mobility of thin film or single crystal transistor. In temperature
dependence, field-effect mobility exhibit gate voltage dependent
activation energy and it shows nonmonotonous temperature
dependence at V$_g$ $<$ -30 V region. This was also observed
irrespective of fabrication method. At nanoscale electrodes we
cannot observe current saturation as in the case of micron scale
FET. The light illumination above absorption edge of
TIPS-pentacene enhanced the saturation current of transistor more
than 4 orders of magnitude and slightly increased its mobility. By
adding off-chip resistor to thin film transistor, we realized
logic circuits such as NOT, NAND, and NOR at room temperature.

\section{Experiment}

TIPS-pentacene (Fig. 1 (a)) was synthesized as described in
elsewhere\cite{Anthony} and it was thermally evaporated with
($\sim$ 70 $^\circ$C) or without substrate heating at base
pressure less than 5 $\times$ 10 $^{-6}$ Torr. After drying of 2
wt. \% toluene solution droplet, single crystalline films are
formed on the surface. Predefined microelectrodes from
photolithography with channel length (L) of 5 $\mu$m $\sim$ 40
$\mu$m and channel width (W) of 400$\mu$m $\sim$ 6 cm were used
for source and drain contacts and heavily doped n-type Si
substrate($\rho$ $<$ 0.002 $\Omega$cm) served as a back gate.
Thermally grown 300 nm SiO$_2$ layer used for gate dielectric
without molecular modification. For the nanoelectrodes
fabrication, we used e-beam lithography and lift-off process.
Electrodes with 200 nm channel length and 2 $\mu$m channel width
were used for nanochannel FET. The thickness of TIPS-pentacene is
50 nm to 800 nm and all the devices are {\it bottom} contact
configuration.

Figure 1 (b) is typical scanning electron microscope (SEM) image
of evaporated TIPS-pentacene film. Its X-ray diffraction data (not
shown) shows clear c-axis orientation as evaporated pentacene
film. Electrical characterization is performed using HP 4155C
semiconductor parameter analyzer.

\section{Results and discussion}
\subsection{Thin film and single crystalline transistor}

Figure 2 is typical output characteristics of TIPS-pentacene FET
at room temperature and inset shows transfer characteristics in
the saturation regime(V$_{DS}$ = -25 V). The gate voltage, V$_g$,
was swept from 0 to -40 V in both cases. This device has room
temperature mobility of 0.02 cm$^2$/Vs and on/off current ratio
higher than 10$^6$. The estimated subthreshold slope is 2.6
V/decade, that is comparable to the pentacene thin film and higher
than some single crystal.\cite{Batlogg} Without any modification
of electrodes and substrate, most of the samples have room
temperature mobility between 0.002 to 0.03 cm$^2$/Vs.

The solution deposited film and evaporated thin film with
substrate heating have slightly higher room temperature mobility
of 0.08 cm$^2$/Vs and 0.05 cm$^2$/Vs and subthreshold swing of 3.8
V/decade and 2.1 V/decade respectively.

Several groups have measured single crystal
FETs\cite{Ramirez,Morpurgo,Park,Podzorov} to get intrinsic
properties of organic FET. We also have used TIPS-pentacene single
crystal as synthesized and put it on top of electrodes under
microscope. To improve contact, transparent polydimethylsiloxane
(PDMS) stamp was laid on the crystal as described in
elsewhere\cite{Park}. The schematic diagram and its optical image
was shown in Fig. 3 (a). The channel length is 100 $\mu$m and
effective channel width from the image is 300 $\mu$m. The output
characteristics (Fig. 3 (b)) also show current enhancement at
negative gate voltage and transfer characteristics (V$_{DS}$ = -
40 V) in the inset show room temperature mobility $\sim$ 0.007
cm$^2$/Vs and subthreshold swing of around 10 V/decade. Compared
to the previous mentioned evaporated or solution deposited FET,
the mobility is decreased and subthreshold swing is increased. A
single crystal FET is very sensitive to surface status, however,
from the optical microscope image the thickness of crystal is not
uniform and the crystal direction is not optimized. Therefore,
this can be improved in uniform and thin crystal with optimized
crystal arrangement.

\subsection{Temperature dependence of mobility}

Figure 4 (a) shows temperature dependence of mobility at different
gate voltages in evaporated film. To avoid persistent
photoconductivity\cite{Brooks2} we kept the sample in dark
condition for a while. At low gate voltage, mobility shows
monotonous temperature dependence with $\mu_{\rm FET}$ $\sim$
exp(-$E_a$/$k_B T$) but at high gate voltage, it reaches maximum
value (T $\sim$ 250 K) and decreases with decrease of temperature.
Unlike the gate-voltage independent mobility of tetracene or
rubrene single crystals\cite{Morpurgo,Podzorov}, mobility of
TIPS-pentacene thin film is gate-voltage dependent and increase
linearly upon gate voltage. This means that induced charge carrier
contribute to the conduction. And trap and grain boundary are
important to this activation type conduction. Inset shows the
change of activation energy depending on the gate voltage. The
activation energy at low gate voltage is E$_a$ $\sim$ 0.27 eV,
which is lower than that of bulk experiment(550
meV).\cite{Brooks2} The activation energy decreased down to 60 meV
with increasing gate voltage. This kind of gate voltage dependent
activation process was widely observed in thin film
pentacene\cite{Brown, Yao}, single crystalline
pentacene\cite{Ramirez,Schoonveld} or
oligothiophene\cite{Schoonveld}. At low temperature region (T $<$
T$_m$), the activation energy is 60 - 80 meV at high gate voltage
(- 40 V $<$ V$_g$ $<$ - 30 V) region.

Typical temperature dependences of FET mobility from different
fabrication processes are shown in Fig. 4 (b). Generally, FETs
fabricated with substrate heating or deposited from solution have
higher room temperature mobility than normally deposited film.
Thin film with substrate heating has activation energy, E$_a$ =
110 - 170 meV with no mobility maximum in measured temperature
range (T$_m$ $>$ 310 K). On the other hand, activation energy of
solution deposited film and thin film evaporated without substrate
heating is 110 meV and 70 meV respectively.  From all the
temperature dependence, sample with higher T$_m$ has high E$_a$,
which means that trap is more dominant in this case therefore
intrinsic transport is suppressed. To explain this nonmonotonous
temperature dependence, several group have adopted following
equation\cite{Morpurgo,Park}

\begin{equation}
\mu \sim T^{-n} {\rm exp}( - \frac{E_t}{k_B T})
\label{equtrap}
\end{equation}

where $E_t$ is the shallow trapping energy and $k_B$ is the
Boltzmann constant.

This nonmonotonous temperature dependence can be understood that
in the high temperature region, microscopic mobility would be
dominant(i.e., intrinsic) and for low temperature region,
thermally activated mobility would be important due to traps. This
was widely observed in tetracene\cite{Morpurgo}, anthracene
\cite{Park}, rubrene\cite{Podzorov}, and thin film
pentacene\cite{Jackson3}.

The activation energy of TIPS-pentacene is larger than thermally
evaporated pentacene thin film transistor (38 meV with room
temperature mobility of 0.3 cm$^2$/Vs)\cite{Jackson3} but less
than precursor-route pentacene thin film\cite{Brown}.

The gate-voltage dependent mobility and nonmonotonous temperature
dependence mean that shallow traps play important role in the
transport of TIPS-pentacene film.

\begin{table}
\caption{Summary of TIPS-pentacene FET characteristics at room
temperature}
\begin{ruledtabular}
\begin{tabular}{ccccc}
Sample&mobility&V$_T$&subthreshold swing&E$_t$\\
&(cm$^2$/Vs)&(V)&(V/decade)&(meV)\\
 \hline
thin film\footnote{without substrate heating.}& 0.02&-14&1.9&70\\
thin film\footnote{substrate heating at 70 $^\circ$C.}& 0.08& -18&2.1&110\\
single crystalline\footnote{solution processed.}&0.05&-12&3.8&170\\
single crystalline\footnote{as synthesized.}&0.007&-15&10.0&na\\
\end{tabular}
\end{ruledtabular}
\end{table}

\subsection{Light illumination}

We illuminated 30 mW He-Ne laser ($\lambda$ = 632.8 nm) to the
evaporated TIPS-pentacene FET, channel length of 5 $\mu$m and
channel width 6 cm. Since the absorption edge of TIPS-pentacene is
around 700 nm, there is nearly 4 orders of current increase upon
light illumination (Fig. 5 (a)). With (Vg= -40V) or without gate
voltage, the current gain, I$_d$$^{light}$/I$_d$, due to light is
$\sim$ 10$^4$. Illumination on the device generate electron-hole
pairs in the channel as well as in bulk TIPS-pentancene. Current
saturation indicates that it is also channel restricted as
observed in photoresponsive polymer FET\cite{Kumar}. Estimated
photoresponsivity, shown in Fig. 5 (a), is around 0.03 A/W and we
can get up to 0.4 A/W in some devices.

We used filter to control optical power. As shown in Fig. 4 (b),
light illumination increases the source-drain current (I$_{DS}$)
and mobility in proportion to optical power. Unlike
semi-logarithmic response of I$_d$ vs intensity in
phototransistor\cite{Lakshmi}, photocurrent increase with light
intensity, I$_{ph}$ $\propto$ P$_{opt}$$^\alpha$. In most
photoconductive polymer films and organic materials, this exponent
$\alpha$ = 1. However, in the case of bimolecular recombination
instead of monomolecular recombination, $\alpha$ = 0.5.\cite{Lee}
The exponent of saturated current, I$_{d,sat}$ vs. optical power
in Fig. 5 (b) is 0.78. When there is carrier traps, immobilized
carrier cannot participate to recombination. This smaller exponent
less than one means that traps play important role.

Generally, the increase of carrier density ($\Delta N$) can be
estimated from $\Delta N$  = C$_i \cdot \Delta V_T /e$, where
C$_i$ is the dielectric capacitance per unit area and $\Delta V_T$
is the shift of the threshold voltage. In this device, $\Delta
V_T$ is around 30 V, hence $\Delta N$ can be estimated to be 1.9
$\times$ 10$^{12}$/cm$^2$. In the case of pentacene\cite{Liang} or
GaN nanowire\cite{Zhou}, light illumination increase only the
carrier concentration and has small effect on the mobility.
TIPS-pentacene shows mobility increase by factor of 5 upon light
illumination.

\subsection{Nanoscale FET}

We reduced the channel length in the submicron range. The
gate-dependent I-V characteristics of 200 nm channel length and 2
$\mu$m channel width FET is shown in Fig. 6. TIPS-pentacene was
thermally evaporated on top of electrodes that was predifined by
e-beam lithography. Unlike the micron scale FET, it doesn't show
any current saturation. This is due to short channel
effect\cite{Sze,Chou}, that source-drain electrodes themselves
make depletion, and space charge limited cunduction (SCLC) is
preventing current saturation. The SCLC follows $I_D$ $\propto$
$V_{DS}^n$, where $n$ depends on the trap concentration and
typically larger than 2. In TIPS-pentacene nanoscale FET, exponent
$n$ is between 2.5 and 3 (shown in the inset of Fig. 6).

Since there is no saturation region in this device, we assume the
lower voltage region to linear regime of transistor for the
estimation of mobility.

\begin{equation}
\frac {\partial{I_D}}{\partial{V_{GS}}} = \frac{W}{L} \mu C_i
V_{DS} \label{equlin}
\end{equation}
where C$_i$ is the gate insulator capacitance per unit area.

The estimated mobility is 0.002 cm$^2$/Vs (at V$_{DS}$ = -2.5 V),
that is similar to lower mobility sample in Fig. 4 (b).

\subsection{Logic gate application}

 We have fabricated inverter (NOT) using this p-type
TIPS-pentacene FET and off-chip resistor as shown in schematic
diagram of Fig. 7 (a). A FET with room temperature mobility of
0.002 cm$^2$/Vs (thickness of 50 nm) and serial resistor of 10
M$\Omega$ were combined for inverter. When the input voltage is
V$_{in}$ $\geq$ 2 V (logical 0), the transistor resistance is
larger than that of series resistor, therefore V$_{out}$ = -10 V
(logical 1). When V$_{in}$ = - 10 V (logical 1), the transistor
resistance is lower than that of series resistor and V$_{out}$ = 0
V (logical 0). The maximum gain defined by dV$_{out}$/dV$_{in}$ is
up to 5.5 (other inverter shows gain up to 8) from the voltage
sweep with driving voltage of -10 V. This value is comparable to
other pentacene inverters\cite{Jackson,Gelinck}. And this gain
value increase as increase V$_d$ as mentioned in Ref.
\cite{Gelinck}. Since all measurements for logic circuits were
done under ambient condition, the device is normally "on" with
positive threshold voltage at room temperature. Some fresh device
show negative threshold voltage but in ambient condition it
shifted to positive as time goes on. That is the reason why we
have to apply positive gate voltage, V$_{in}$ to "off" the FET .
In actual application, this need additional level-shifting
circuit. To avoid this complication, it's important to reduce air
exposure and {\it light} illumination.

By adding one more FET, NAND (Fig. 7 (b)) and NOR (Fig. 7 (c))
logic were constructed. Since the FET is "on" at zero gate voltage
(i.e. positive threshold voltage), we applied positive gate
voltage to "off" the transistor. In this operation, we used "+10
V" for off state. Due to mismatch of current and mobility of two
transistors, there is shift of offset voltage in "off" state.\\

\section{Summary}

In summary, we measured the temperature dependence of mobility of
functionalized pentacene(TIPS-pentacene). It shows gate-voltage
dependent activation energy that is in the 60 - 170 meV at high
gate voltage (V$_g$ $<$ -30 V) depending on the fabrication
procedure. Temperature dependence can be ascribed to shallow
trapping of charge carriers. Light illumination increase the
mobility but it is photoconductive response. Logic gate circuit
were also demonstrated. Inverter has gain upto 5.5 that is
comparable to other pentacene device and NAND and
NOR logic gates were also presented.\\

This work is supported by NSF-DMR-0203532. JGP thanks to MARTECH
for using their equipments.

\newpage

\begin{figure}[b]
\includegraphics*[bb=60 277 561 518]{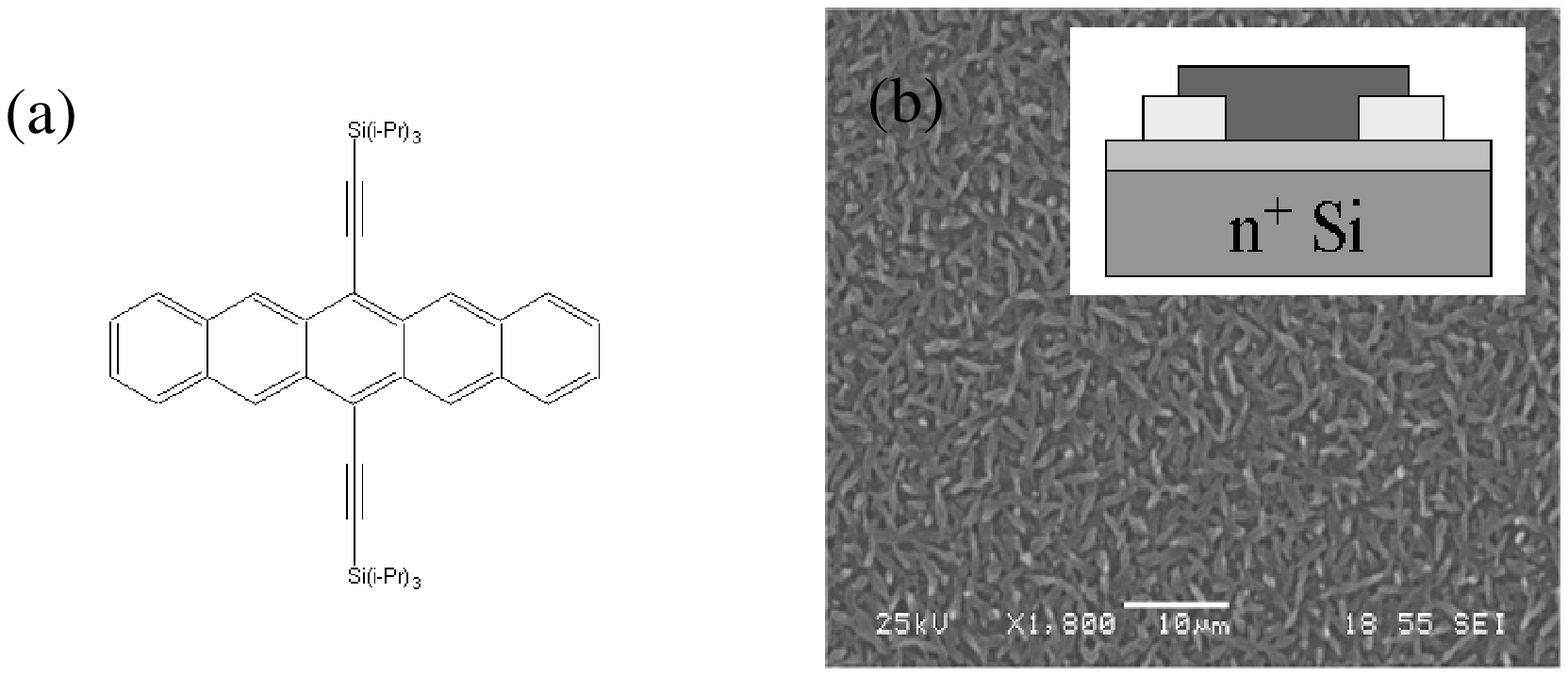}
\caption{(a) Chemical structure of TIPS-pentacene. (b) SEM image
of thermally evaporated TIPS-pentacene} \label{fig1}
\end{figure}

\newpage

\begin{figure}[b]
\includegraphics*[bb=54 225 557 583]{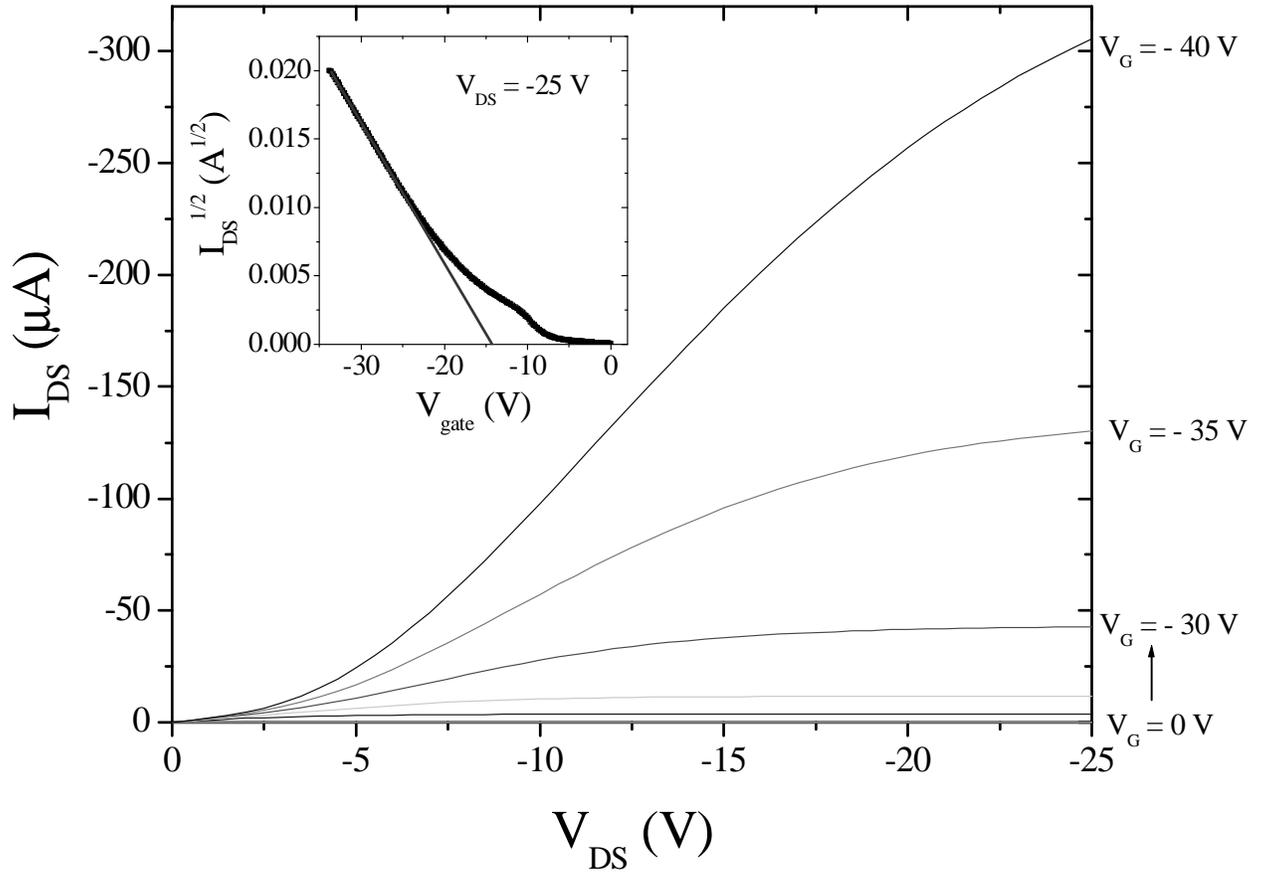}
\caption{Typical output characteristics of TIPS-pentacene
transistor with channel length of 5 $\mu$m and channel width of
6cm in "bottom" contact configuration. From the transfer
characteristics, estimated subthreshold slope is 2.6 V/decade and
mobility is 0.02 cm$^2$/Vs with threshold voltage of V$_T$ = - 14
V.} \label{fig2}
\end{figure}
\newpage

\begin{figure}[t]
\includegraphics*[bb=51 90 554 687]{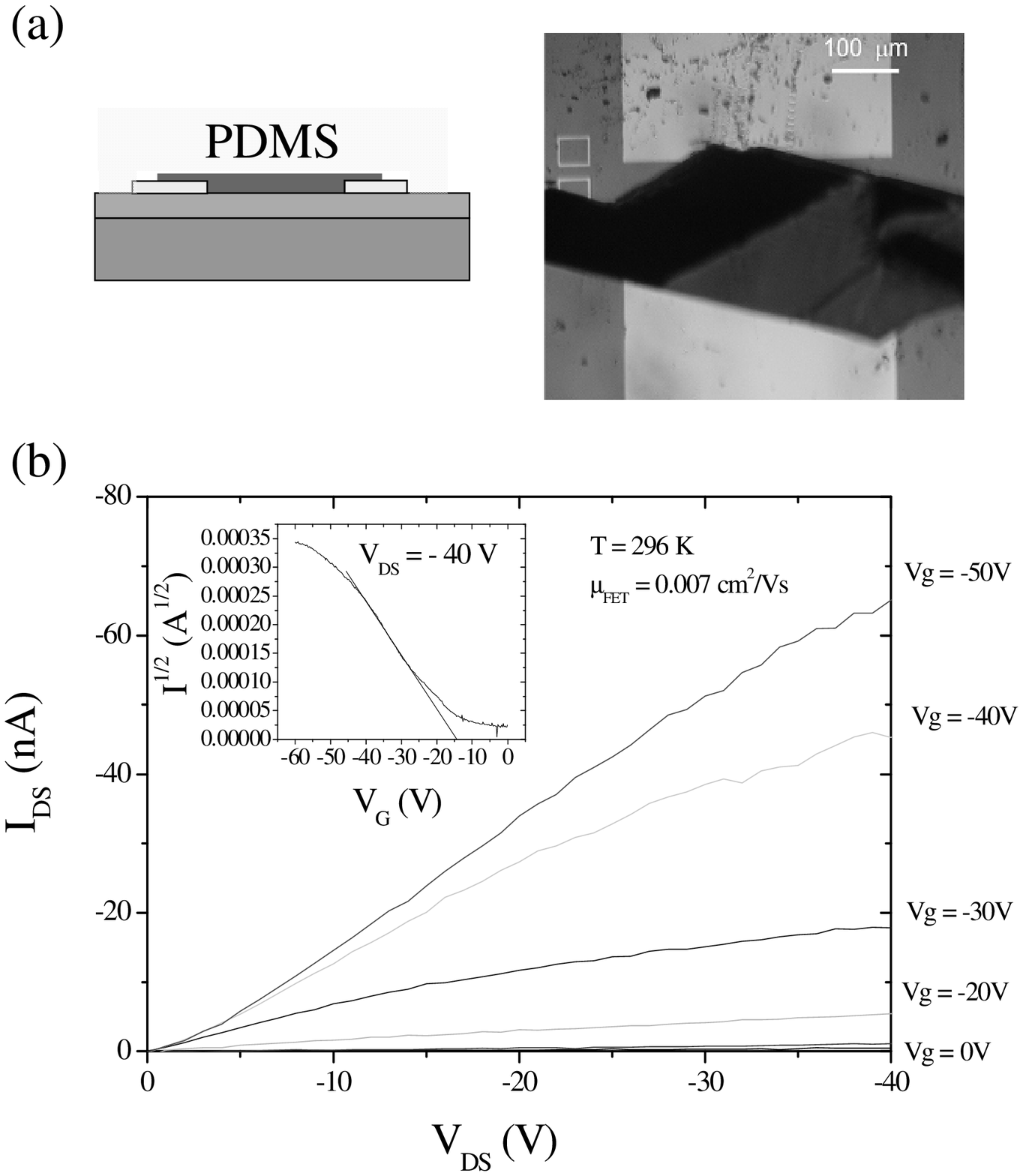}
\caption{(a) Schematic diagram for single crystal FET pressed by
PDMS to improve contact and its optical microscope image. (b)
Output characteristics measured at room temperature. Inset shows
transfer characteristics at V$_{DS}$ = -40 V.} \label{fig3}
\end{figure}

\begin{figure}[t]
\includegraphics*[bb=71 55 507 716, width=14cm]{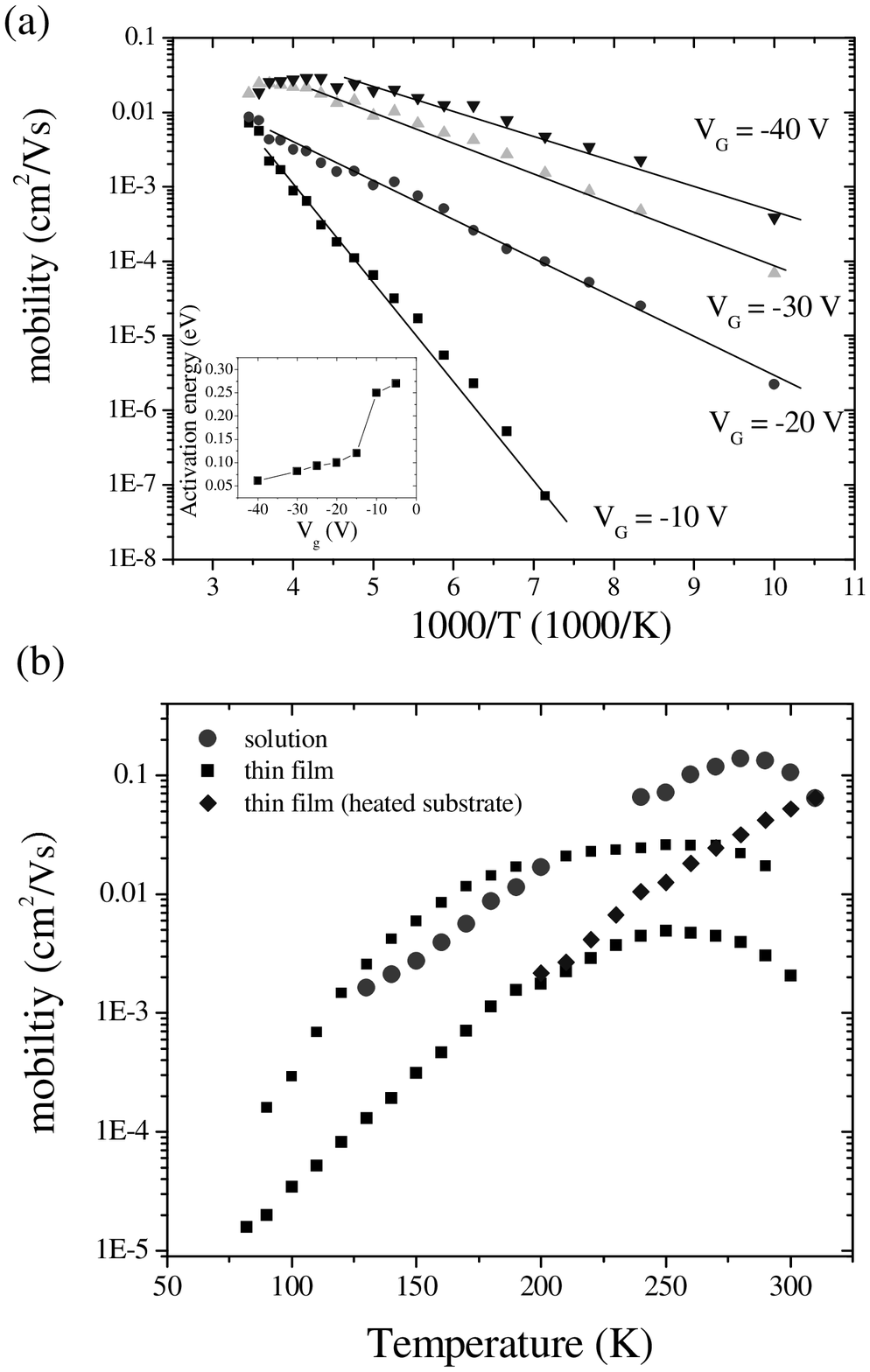}
\caption{(a) Temperature dependence of mobility at different gate
voltage. TIPS-pentacene was deposited without substrate heating.
Inset shows the change of activation energy upon gate voltage. It
decreases from 0.27 eV (V$_g$ = - 5V) to 60 meV (V$_g$ = -40 V).
(b) Temperature dependence of mobility from different fabrication
processes such as evaporated thin film with($\blacklozenge$) or
without substrate heating($\blacksquare$) or solution
deposition($\bullet$).} \label{fig4}
\end{figure}
\newpage

\begin{figure}[tbp]
\includegraphics*[bb=106 160 510 731]{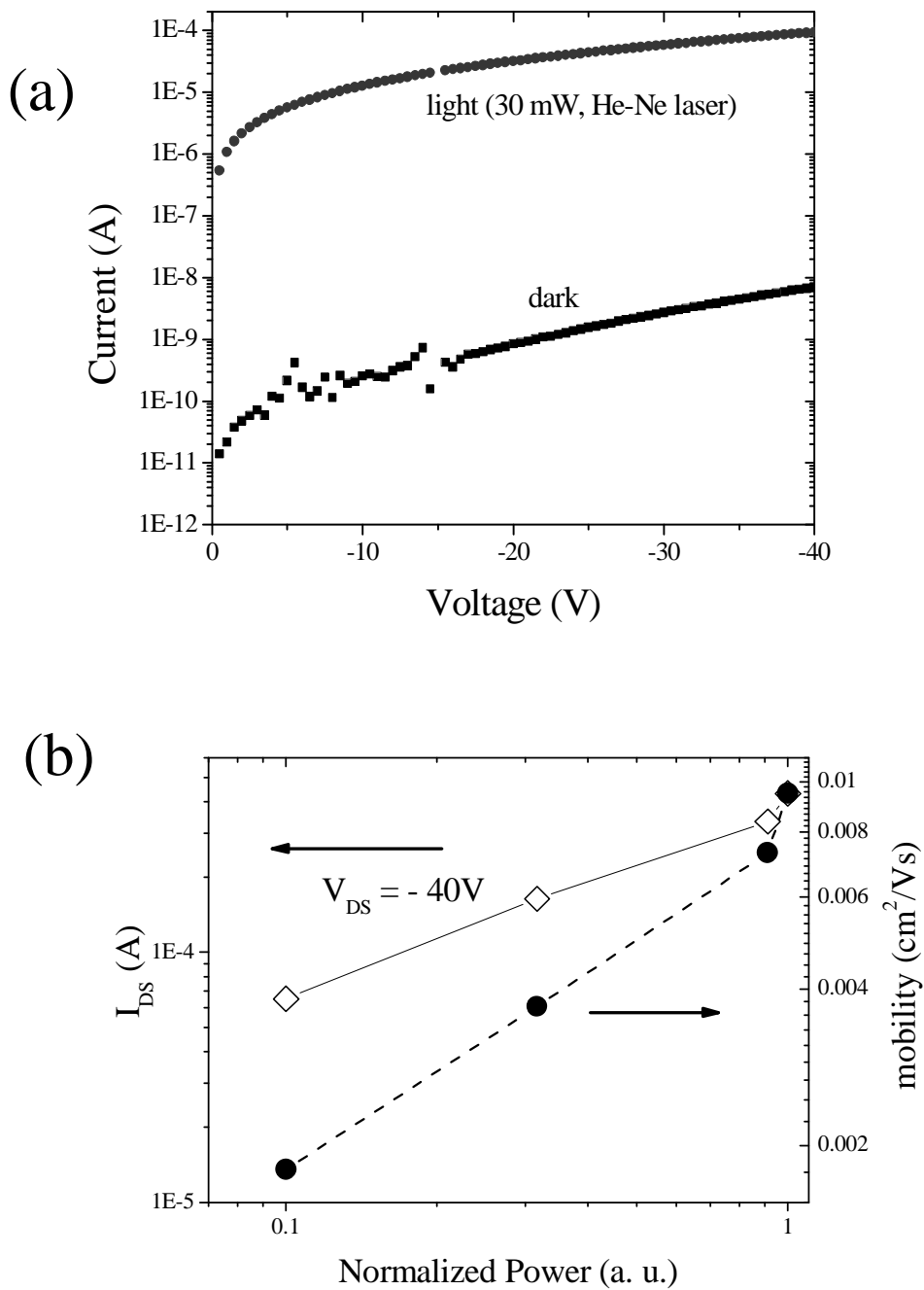}
\caption{(a) Change of I-V characteristics of TIPS-pentacene under
the illumination of He-Ne laser (632.8 nm, 30 mW) without gate
voltage. (b) I$_{DS}$ (at V$_{DS}$ = -40 V, V$_G$= 0 V) and
field-effect mobility as a function of illuminated light power at
ambient condition.} \label{fig5}
\end{figure}
\newpage

\begin{figure}[tbp]
\includegraphics*[bb=60 224 511 595]{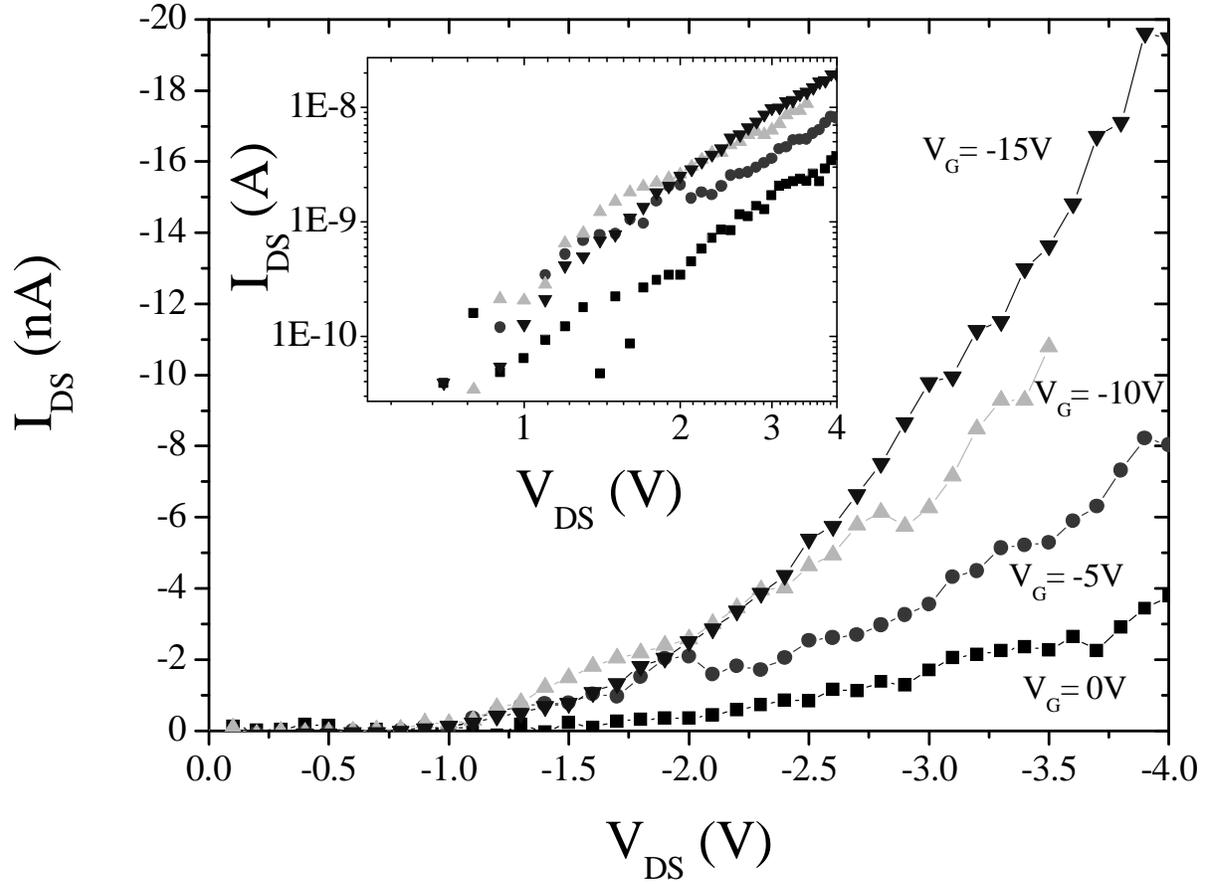}
\caption{Gate effect of nanochannel FET of TIPS-pentacene. There
is no current saturation in nanoscale device due to short channel
effect. Inset shows log-log plot of output characteristics, which
follow SCLC.} \label{fig6}
\end{figure}

\newpage
\begin{figure}[tbp]
\includegraphics*[bb=45 66 532 778, width=14cm]{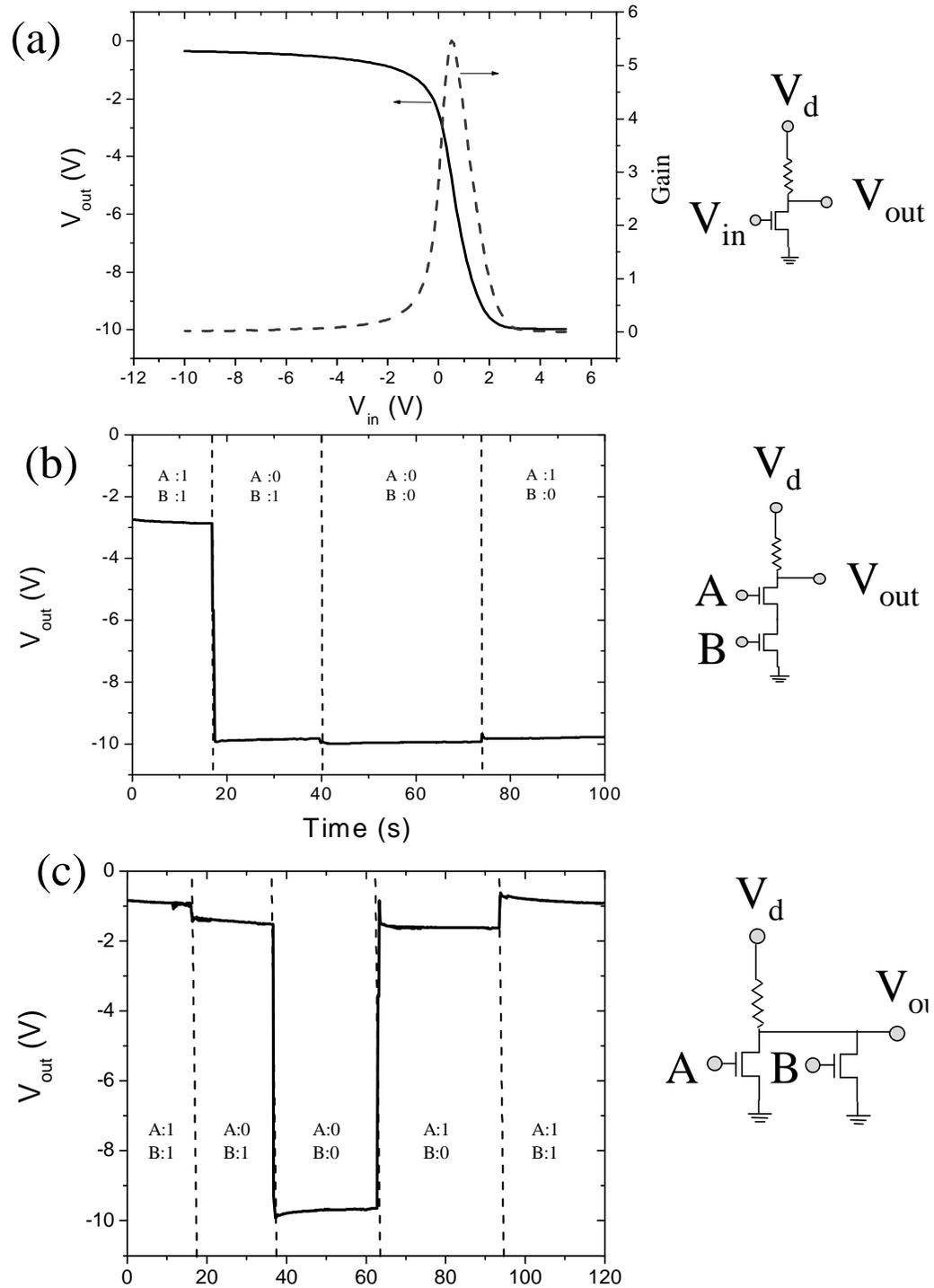}
\caption{(a) inverter, (b) NAND gate logic, and (c) NOR gate logic
circuit made from thin film TIPS-pentacene with room temperature
mobility of 0.003 cm$^2$/Vs. Loading voltage (V$_d$) of -10 V and
off-chip resistor of 10 M$\Omega$ are used in the circuit.}
\label{fig7}
\end{figure}

\end{document}